\documentstyle{cupconf}
\input{psfig}

\ifoldfss
\else
  \ifnfssone
    \newmathalphabet{\mathit}
      \addtoversion{normal}{\mathit}{cmr}{m}{it}
      \addtoversion{bold}{\mathit}{cmr}{bx}{it}
    \newmathalphabet{\mathcal}
      \addtoversion{normal}{\mathcal}{cmsy}{m}{n}
    \else
    \ifnfsstwo
    \fi
  \fi
\fi

\def\mb#1{\setbox0=\hbox{$#1$}\kern-.025em\copy0\kern-\wd0
\kern-0.05em\copy0\kern-\wd0\kern-.025em\raise.0233em\box0}
\def\hexnumber#1{\ifcase#1 0\or1\or2\or3\or4\or5\or6\or7\or8\or9\or
 A\or B\or C\or D\or E\or F\fi }

\makeatletter
\ifx\CUP@mtlplain@loaded\undefined
\else
\fi
\makeatother
 \makeatletter
 \ifx\CUP@mtlplain@loaded\undefined
   \font\tenbmi=cmmib10 at 10pt
   \font\sevenbmi=cmmib10 at 7pt
   \font\fivebmi=cmmib10 at 5pt

   \newfam\bmifam
   \textfont\bmifam=\tenbmi
   \scriptfont\bmifam=\sevenbmi
   \scriptscriptfont\bmifam=\fivebmi
   
 \fi
 \makeatother
\ifnfsstwo

\fi
\ifnfssone

\fi
\ifoldfss

\fi

\mathchardef\varLambda="0103

\makeatletter
\ifx\CUP@mtlplain@loaded\undefined
\else
\fi
\makeatother
\makeatletter
\ifx\CUP@mtlplain@loaded\undefined
  \font\tenbms=cmbsy10
  \font\sevenbms=cmbsy10 at 7pt
  \font\fivebms=cmbsy10 at 5pt
  \newfam\bmsfam
  \textfont\bmsfam=\tenbms
  \scriptfont\bmsfam=\sevenbms
  \scriptscriptfont\bmsfam=\fivebms

  \edef\bsy@{\hexnumber\bmsfam}
  \mathchardef\bnabla="0\bsy@72
\fi
\makeatother
\title[Spotted disks]{Spotted disks}

\author[A. Bracco {\it et al.\/}]%
{A.\ns B\ls R\ls A\ls C\ls C\ls O$^1$,
\ns
A.\ns P\ls R\ls O\ls V\ls E\ls N\ls Z\ls A\ls L\ls E$^{1,2}$,\ns\\
E.\ns A.\ns S\ls P\ls I\ls E\ls G\ls E\ls L$^3$
\and \ns P.\ns Y\ls E\ls C\ls K\ls O$^4$}

\affiliation{$^1$Istituto di Cosmogeofisica, Corso Fiume 4, I-10133
Torino, Italy\\[\affilskip]
$^2$ JILA, University of Colorado, Box 440, Boulder, CO 80309-0440,
USA\\[\affilskip]
$^3$ Dept. of Astronomy, Columbia University, New York, NY 10027, USA
\\[\affilskip]
$^4$ Dept. of Physics, University of Florida, Gainesville, FL, USA
}

\begin{document}
\ifnfssone
\else
  \ifnfsstwo
  \else
    \ifoldfss
      \let\mathcal\cal
      \let\mathrm\rm
      \let\mathsf\sf
    \fi
  \fi
\fi

\maketitle

\begin{abstract}
Rotating, turbulent cosmic fluids are generally pervaded by
coherent structures such as vortices and magnetic flux tubes.
The formation of such structures is a robust property of rotating
turbulence as has been confirmed in computer simulations and laboratory
experiments.  We defend here the notion that accretion disks share
this feature of rotating cosmic bodies.
In particular, we show that the intense shears of Keplerian flows do not
inhibit the formation of vortices.  Given suitable initial disturbances
and high enough Reynolds numbers, long-lived vortices form in Keplerian
shear flows and analogous magnetic structures form in magnetized disks.
The formation of the structures reported here should have significant
consequences for the transport properties of disks and for the
observed properties of hot disks.

{\bf NOTE: figures here have been bitmapped to conform to xxx.lanl standards, please
contact A. Bracoo or P. Yecko for original figures}

\end{abstract}

\firstsection 
\section{Introduction}

Whenever it has been possible to observe rotating, turbulent fluids with
good resolution, it has been seen that individual, intense
vortices form (Bengston \& Lighthill 1982;
Hopfinger {\it et al.} 1982; Dowling and Spiegel 1990).
In each case where the
medium is a good conductor, as in the solar convection zone, the vortices
are instead magnetic flux tubes but the guiding principle remains the
same: vector fields, whether vorticity or magnetic fields, are amplifed
and concentrated by the combined effects of rotation and turbulence.

It seems natural that one should apply this principle to the
theory of accretion disks (Dowling \& Spiegel 1990).  However, as
critics of this idea have objected, accretion disks are exceptional among
rotating turbulent objects in the strong shears that these bodies are
believed to possess and this shear might lead to rapid destruction of any
structures that
tend to form.  This objection is here overruled by a
large number of simulations.  Naturally, this does not prove that coherent
structures must form on disks, but it does strengthen the argument that
disks are likely to follow the norm of rotating, turbulent bodies.
We may point also to
numerical results suggesting that vortices
form on disks (Hunter \& Horak 1983) as well as experimental work
(Nezlin \& Snezhkin 1993).

Even if it is granted that vortices and magnetic flux tubes can form
on accretion disks, one must naturally ask whether this matters for
processes
such as transport rates or for observational
properties.  For both these issues, we suggest that there can indeed
be an important role of vortices.  As we shall explain in the fluid
dynamical review given below, there is an important property of rapidly
rotating fluids called the potential vorticity that is approximately
conserved following the motion of the fluid (Ghil \& Childress 1987).
When a vortex is perturbed (for example, by interaction with other
vortices) it will be set to oscillating and will radiate fluid dynamical
waves, especially Rossby waves (Llewellyn Smith 1996).  This radiation
represents a loss of specific angular momentum and, to conserve potential
vorticity, the vortex must move to a location with a different mean
angular velocity.  This results in radial angular momentum transport.
Moreover, coherent vortices trap passive tracers for long times
(Elhmaidi {\it et al.} 1993; Babiano {\it et al.} 1994) and are
responsible for much of the transport of passive and active constituents
in rotation-dominated flows, see Provenzale {\it et al.} (1997) for a
general review and Tanga {\it et al.} (1996) for an application to
the early solar nebula.

It may be true that the fluid dynamics of disks, particularly of the
turbulence in them, is so poorly understood that one should not worry
about such effects, unless they threaten to become dominant.  On this
issue, unfortunately, we remain uncertain.  But the possible observational
effects of vortices (and flux tubes) do seem to us to have a real
importance.

The centrifugal force or magnetic pressure produced by vortices
produce local extrema in the total pressure (gas plus radiation).
Hence, these structures can act either like holes in the
disks or thickenings according to whether the vortices are
cyclonic (having the sense of the local rotation) or anticyclonic.
The holes may even go all the way through the disk in some cases, allowing
the rapid escape of hard radiation from within the disk.  Such radiation,
coming from within an optically thick hot disk, can produce spots on disks
that may affect the observed spectra of AGNs (Abramowicz {\it et al.}
1991).

In this paper, we shall not enter into the discussion of these
motivating topics, but rather shall stick to the immediate fluid dynamical
issues.  We shall review some of the fluid equations of the disk problem
and report on simulations that suggest that vortices form in the presence
of intense shears.  We realize that this suggestion has not gained
general acceptance among specialists in disk theory, but are encouraged
by the recent softening of their attitude toward the issue of vortex
formation on disks.  Thus, several years ago, when we suggested at the
cataclysmic variable meeting in Eilat that disks would have spots, with
the consequence that phenomena associated with the solar cycle might be
expected around disks, including hot coronas and long term variations,
the summarizer of that meeting disparaged these suggestions. At the
meeting in Iceland, at which the present version was reported, the
summarizer merely ignored our remarks and did not mention them at all.
Given this apparent warming of the climate of opinion, we are encouraged
in this case to report our findings here.   These strongly suggest that
disks are not so different from other rotating cosmic bodies and that they
produce coherent structures that may play a role in their large scale
properties.

\section{Thin-Layer Theory}
Our discussion focuses on fluid dynamical issues with little
attention paid to the thermal aspects of accretion flows.
The Navier-Stokes and continuity equations then
govern the
fluid fields, velocity ${\bf v}=(u,v,w)$ and density $\rho$, which
depend on position and time.  These equations are
\begin{equation}
{{ D {\bf v}}\over {Dt}}=-{1\over \rho}\nabla p
-\nabla \Phi +{\bf D}_{\bf v}
\label{NS}
\end{equation}
and
\begin{equation}
{ {D \rho}\over {Dt}}+\rho \nabla \cdot {\bf v}=0
\label{cont}
\end{equation}
where
$D/Dt=\partial/\partial t + {\bf v}\cdot \nabla$ is the material
derivative, $p$ is pressure and $\Phi$ is the potential of the (imposed)
gravitational field.  We shall be concerned with flow at some distance
from the central object and will assume that $\Phi=-G M/R$, where
$R$ is the distance from that object
and $M$ is its mass.
The last term in (\ref{NS})
is the viscous force per unit mass, which, for constant viscosity, is
\begin{equation}
{\bf D}_{\bf v} = \nu[\nabla^2 {\bf v} + {1\over 3} \nabla
(\nabla \cdot {\bf v})
] \ .
\label{visc}
\end{equation}

In discussions of rotating and turbulent fluids the vorticity
field, ${\mb
{\omega}}=\nabla \times {\bf v}$, proves to be a very
important quantity.  We obtain an equation for it by taking the curl
of eq.(\ref{NS}); this is
\begin{equation}
{D{
\mb
{\omega}}\over Dt} + {
\mb
{\omega}}(\nabla \cdot {\bf v})
= {
\mb
{\omega}} \cdot \nabla {\bf v} + {\nabla \rho \times \nabla p \over
\rho^2} + {\bf D}_{
\mb
{\omega}}
\label{vorte}
\end{equation}
where ${\bf D}_{
\mb
{\omega}}=\nabla \times {\bf D}_{\bf v}$.
We see that
if the surfaces of constant $p$ and $\rho$ are not coincident, vorticity
is generated by the so-called baroclinic term.  We shall not include this
effect here.  Instead, we assume that the fluid is barotropic, which means
that the pressure is a function only of the density.  A familiar example
of such a relation is the adiabatic gas law
\begin{equation}
p=K \rho^\gamma
\label{poly}
\end{equation}
where $K$ and $\gamma$ are constants.  This holds for a perfect
gas when the specific entropy, $S=c_v log(p/\rho^\gamma)$,
is constant in space and time.

In a thin layer of fluid, there is not much horizontal vorticity
and we are concerned mainly with the component of ${
\mb
{\omega}}$
parallel to the rotation axis, which, in the case of a disk with rough
cylindrical symmetry, we associate with the $z$-axis, or vertical
direction.  We call this component of vorticity, the vertical component,
$\zeta$.  When ${
\mb
{\omega}} \approx (0,0,\zeta)$,
the fluid velocity is
approximately horizontal.  This is the thin-layer approximation
that
is used in the study of shallow layers in geophysical and
planetary fluid dynamics.  The idea is that
if the vertical velocity were not small
the layer would not be thin.
Then we may approximate
(\ref{vorte}) as
\begin{equation}
{{D\zeta}\over {Dt}} + \zeta \nabla \cdot {\bf u} = D_\zeta
\label{vorticity}
\end{equation}
where ${\bf u}=(u,v)$ is the horizontal velocity (in the plane
of the disk) and $D_\zeta$ is the $z$-component of
${\bf D}_{
\mb
{\omega}}$.

There are two ways to derive thin-layer equations in a more formal
and systematic way other than the appeal to physical intuition
that we make here.  One is by integrating over $z$,
with certain assumptions on the vertical structures of the fields
and the other is to make expansions in terms of small thickness
(Ghil \& Childress 1987; Qian {\it et al.} 1991).
These approaches work as well on compressible as on incompressible
fluids, as long as the motions remain reasonably subsonic.  The
resulting equations are those of a compressible fluid, but it is the
surface density that is varying strongly, whether or not the real density
varies.  We see this in the integrated continuity equation.

To reduce (\ref{cont}) to an equation for the surface density we consider
here only the case where the layer is symmetric under the transformation
$z\rightarrow -z$.  We assume that the disk has a surface on which the
density goes to zero and that this is given by a height function $h$ as
$z=\pm h(x,y,t)$. The surface density is then defined as
\begin{equation}
\sigma(x,y,t) = \int_{-h}^h \rho(x,y,z,t) dz.
\label{sigma}
\end{equation}

For barotropic, axisymmetric stationary disks, it may be shown that the
velocity is independent of $z$.  Naturally, this will not be exactly right
for turbulent, time-dependent disks without axisymmetry, but if
the violations are mild, we may use this picture in first approximation.
Then, if we integrate the continuity equation over the depth of the
disk, we obtain
\begin{equation}
{{D\sigma}\over {Dt}} + \sigma \nabla\cdot {\bf u}=0.
\label{surf}
\end{equation}
We see from the structure of this equation that the dynamical
description of a thin disk is that of a compressible, two-dimensional
fluid with density $\sigma$.  But $\sigma$ is really a
surface density and its variations can reflect either real density
variations or thickness variations.  Since the thin-layer description is
the same for gases and liquids, apart from coefficients in the equations,
the term density wave has therefore to be used with caution since the real
density variations associated with variations of $\sigma$ need not be
large.  In particular, the waves associated with gravitational instability
in thin layers are gravity waves and not sound waves (Qian \& Spiegel
1994).

Combining equations (\ref{vorticity}) and (\ref{surf}) we
obtain
\begin{equation}
{{ D q}\over {Dt}} = D_q
\label{shallow}
\end{equation}
where
\begin{equation}
q(x,y,t) = {{\zeta(x,y,t) }\over {\sigma(x,y,t)}}
\label{pote}
\end{equation}
is the potential vorticity in the shallow-layer approximation
and the dissipation term is $D_q=D_\zeta/\sigma$.

In the absence of dissipation ($D_q=0$) in (\ref {shallow}) the
potential vorticity becomes a material invariant, that is, $Dq/Dt=0$.
This condition is central to many discussions of planetary and geophysical
fluid dynamics (Pedlosky 1987).  The approach in those subjects is to use
an approximation to provide the velocity in the $D/Dt$ of this equation.
The approximation usually used is geostrophy, in which the Coriolis force
is exactly balanced by the pressure gradient.  In the presence of the
intense shear of accretion disks, this approximation is less useful than
in those other subjects and we shall, in this section, adopt another
approach.

In order to concentrate on the behaviour of vorticity and the effects of
shear, we shall omit the variations in surface density, both
globally and locally.  Then, by requiring that $\sigma$ be constant
we find that $\nabla \cdot {\bf u}=0$ where ${\bf u}\equiv (u,v)$ is the
(horizontal) velocity in the disk and ${\bf r}=(x,y)$ are the horizontal
coordinates.   Evidently, the resulting description is formally equivalent
to two-dimensional incompressible hydrodynamics and we may introduce a
stream function $\psi$ such that
\begin{equation}
(u,v)=\left( -{\partial \psi \over \partial y} \ , {\partial \psi \over
\partial x}
\right) \ . \label{psi}
\end{equation}
Then the (total) vertical vorticity,
\begin{equation}
\zeta=\nabla^2\psi
\
,
\label{zeta}
\end{equation}
is the potential vorticity and we use equations (\ref{vorticity}),
which can be written in the form
\begin{equation}
{{\partial \zeta} \over {\partial t}} + [\psi,\zeta] = D_\zeta
\label{2d}
\end{equation}
where $[\psi,\zeta]=(\partial_x) \psi \partial_y \zeta -  (\partial_x
\zeta) \partial_y \psi$.

For a solenoidal velocity, the viscosity term simplifies and we have
followed standard practice in fluid dynamics by considering a variety of
representations of it, all of which may be written as
\begin{equation}
D_\zeta = (-1)^{p-1} \nu_p \nabla^{2p} \zeta \
\label{dissip}
\end{equation}
with $\nu_p$ taken constant.  The case $p=1$ corresponds to the
conventional Newtonian viscosity but values of $p > 1$, the so-called
hyperviscous cases, have also been used.  Hyperviscosity is a
device used in turbulence simulations to represent the effects of eddy
viscosity, see e.g. McWilliams (1984).
Its advantage is that it cuts off the viscous tail of the
spectrum rapidly so that high spatial resolution is more easily achieved
in a given computational time.

With this specification of $D_\zeta$, we first use
(\ref{zeta})-
(\ref{2d}) to study the formation of coherent structures.  This is
physically reasonable when (a) the
local
vorticity
of
the rotation
of the disk is larger than the vorticity
of
perturbations of
the basic state and (b) the dynamics takes place on a scale smaller than
that over which the disk thickness varies appreciably. Since vorticity
has the dimensions of inverse time, (a) is valid when the dynamics of the
perturbed flow is slower than the typical rotation time scale of the whole
disk at the given radius. If we indicate by $\tilde \zeta$ a typical
vorticity perturbation, and by $\Omega$ the mean angular velocity at the
given radius, we expect that (\ref{2d}) is valid in the limit $\tilde
\zeta/f \ll 1$ where $f=2\Omega$
is the local Coriolis frequency.

Since the typical order of magnitude of $\tilde \zeta$ may be written as
$[\tilde \zeta]=U/L$ where $U$ and $L$ are a typical velocity and
length scale of the perturbation, one can express the above constraint
by the requirement $Ro=U/(f L) \ll 1$, where $Ro$ is called
the Rossby number. The smallness of the Rossby number is a
requirement for approximating the shallow-layer equation
(\ref{shallow}) by the quasi-geostrophic vorticity equation
(Pedlosky 1987).  Since $\Omega=\Omega(|{\bf r}|)$, the Rossby
number depends on the radial coordinate of the disk.

In the absence of dissipation ($D_\zeta = 0$), eq.(\ref{2d}) admits an
infinite number of conserved quantities,
two of which
are quadratic
invariants. These are the kinetic energy $E = 1/2 \int{(\nabla
\psi)^2 dxdy}$ and the enstrophy $Z=1/2 \int{(\nabla^2 \psi)^2 dxdy}$.
The conservation of these two quantities induces a direct
cascade of enstrophy from large to small scales and an inverse (from
small to large scales) cascade of kinetic energy
(Charney 1971; Rhines 1979; Kraichnan \& Montgomery 1980).
This is different
from the more familiar direct cascade of energy in
three-dimensional
turbulence, where the energy flows from large to small scales.

For freely decaying
($D_\zeta \ne 0$) barotropic turbulence in shear-free
environments, intense long-lived vorticity concentrations are observed to
form after an energy dissipation time.  These coherent objects
are characterized by a broad distribution of size and circulation and
they contain most of the energy and the enstrophy of the system
(McWilliams 1984; McWilliams 1990).
The question addressed here is whether the strong Keplerian shears of
accretion disks modify this well-documented behaviour.

\section{Simulations with Keplerian Shear}

In the case of the Keplerian disk, where the shear and differential
rotation are quite strong, the question of vortex formation and survival
remains open and we address it in this section.
By using the same formalism as in previous simulations of vortices,
we may isolate the effect of the shear, which favours the propagation of a
class of dispersive waves, called Rossby waves
(Pedlosky 1987),
that do not occur in
plane layers with rigid background rotation. Rossby waves transport energy
away from a source of perturbation and so can inhibit the formation of
large-scale vortices. Analytical studies and numerical simulations have
shown that, in the so-called $\beta$-plane --- a slab with linear
differential rotation, $2\Omega=f(y)=f_0+\beta y$ --- coherent vortices
form and dominate the dynamics on small scales, while Rossby waves
inhibit the formation of vortices at large scales.  The larger is
$\beta$, the smaller is the scale below which vortices can survive
(Cho \& Polvani 1996a; 1996b).

Here, we report simulations based on a standard pseudo-spectral
code with 2/3 dealiasing and resolution
$512\times 512$
grid points
in a square box with periodic boundary conditions. After checking that
the value of $p$ had no significant effect on the outcome, if sufficient
running time was allowed, we have used the hyperviscosity with $p=2$
for most of this work.

We start each calculation with an initial condition corresponding to
a perturbed Keplerian disk, namely
\begin{equation}
\zeta({\bf r},t=0) = \Omega(r) + \tilde \zeta({\bf r},0)
\label{ini}
\end{equation}
where
$r=|{\bf r}|$, $\Omega(r)=K r^{-3/2}$ is the Keplerian vorticity
profile, $K$ is a constant depending on the
mass of the central object, and $\tilde \zeta({\bf r},0)$ is the
initial vorticity
perturbation which is superposed on the Keplerian profile.
The center of the disk is placed at the center of
the simulation box
at
${\bf r}=(0,0)$ and
the initial Keplerian profile has been smoothed
at the center to eliminate the vorticity singularity in ${\bf r}=0$.

In this section, we use dimensionless variables.
The space scale is fixed by defining the simulation box to have
size $2\pi$.  The time scale is fixed by the imposed values of the energy
and of the enstrophy. In the simulations discussed below,
we have chosen the initial kinetic energy of the unperturbed
Keplerian disk to be $E_{kep}=2.0$.  This gives a Keplerian initial
enstrophy $Z_{kep}\approx 6$.  At $r=\pi/4$, the local Keplerian
vorticity is $\Omega(\pi/4)\approx 6$, which gives a typical rotation time
$T_{kep}(r=\pi/4)=2\pi/\Omega \approx 1$.  The simulations have in
general
been run up to a total time $T=20$, which means $\approx 20$ rotations at
$r=\pi/4$.  The initial energy of the perturbation, $\tilde E$, has then
been fixed in the range $ 0 \le \tilde E/ E_{kep} \le 0.1$.  The eddy
viscosity coefficient has been fixed at the value $\nu_2=5\cdot 10^{-8}$.

Simulations starting with a pure Keplerian profile
(i.e., $\tilde E = 0$) have indicated that this profile is
stable to small disturbances.  The perturbations involved in normal
numerical procedures produce no instability and the disk shear slowly
decays due to dissipation. The decay time is extremely long and the shear
survives for hundreds of rotation times $T_{kep}(\pi/4)$.

In the course of this study we have considered three types of
initial perturbations $\tilde \zeta({\bf r},0)$: a single vortex structure,
radially localized waves with various azimuthal wavenumbers, and
a random initial perturbation field. All of these initial
conditions produce analogous results. In the following, we present the
evolution of the disk resulting from a random initial perturbation
field $\tilde \zeta$, as this represents the most physically
plausible type of perturbation.  The initial perturbation has been
generated as a narrow-band random vorticity field with energy spectrum
\begin{equation}
E(k)=E_0{ { k^n} \over {[(m/n)k_0+k]^{m+n}}}
\label{spectrum}
\end{equation}
where $E_0$ is a normalization factor; we used $k_0=10$,
$n=5$ and $m=30$.
The Fourier phases are randomly distributed between $0$ and $2\pi$. In
order to avoid unrealistic periodicity effects, the perturbation has been
set to zero for $r > 2\pi/3$, such that no perturbation is present at the
edges of the simulation box.

For small perturbation energies, the vorticity perturbations are
sheared away by the Keplerian flow, and the disk quickly returns to
an unperturbed, slowly decaying Keplerian profile.
This behavior is consistent with the linear stability of Keplerian
profiles.

For large perturbation energies, the initial random vorticity
field self-organizes into coherent vortices, similarly to
what happens in non-shearing conditions.  We refer to the sense
of the basic Keplerian rotation as cyclonic (counterclockwise in the
figures shown here).   In the presence of the basic Keplerian shear
cyclonic
disturbances are rapidly sheared away, and are not able to organize
themselves into individual vortices. By contrast, initial anticyclonic
(negative vorticity) perturbations grow and form
coherent
vortices.  Once formed, these anticyclonic vortices merge with each
other, and generate larger vortices. This growth by merger is halted
by the Keplerian shear, that, analogously to what happens for turbulence
on the $\beta$-plane, sets an upper limit to the size of the surviving
vortices.  The anticyclonic vortices are coherent in the sense that they
live much longer than the typical eddy turnover time of the perturbation,
$\tilde T=\tilde Z^{-1/2}$ where $\tilde Z$ is the perturbation enstrophy.

Figures 1, 2 show the vorticity 
field for a simulation, in which the initial
perturbation energy was normalized so that $\tilde E/E_{kep}\approx
0.005$.  The vorticity is shown at the times $t=5$ (figure 1) 
and $t=10$ (figure 2).
Figure 3 shows a section of the vorticity
field at time $t=0$ (the initial condition) and at time $t=5$.
The
average vorticity profile remains Keplerian for the whole
simulation
and the rotation curves show bumps like those seen on rotation
curves of galaxies.

\begin{figure}

 \centerline{\psfig{figure=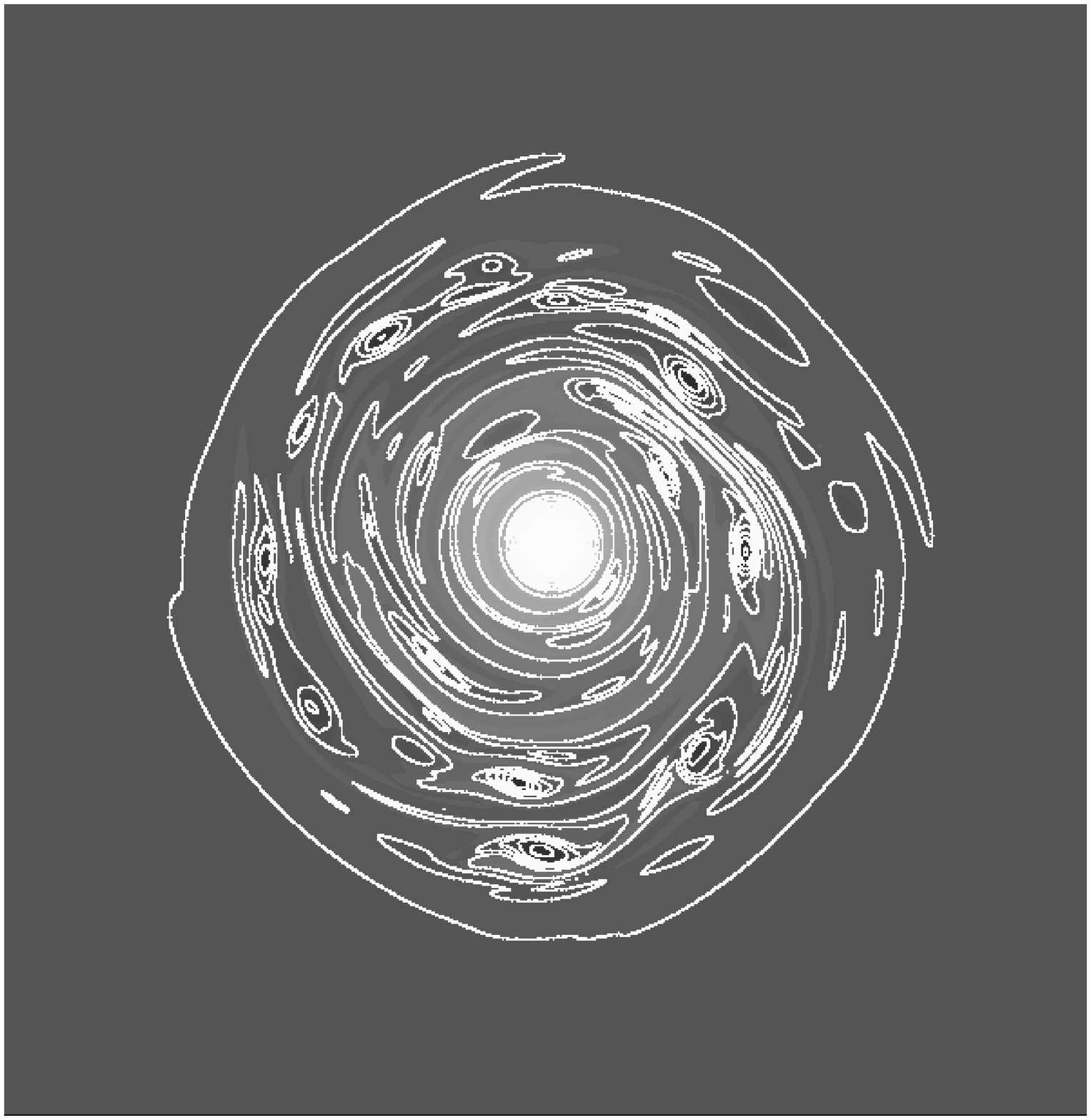,height=7cm}\psfig{figure=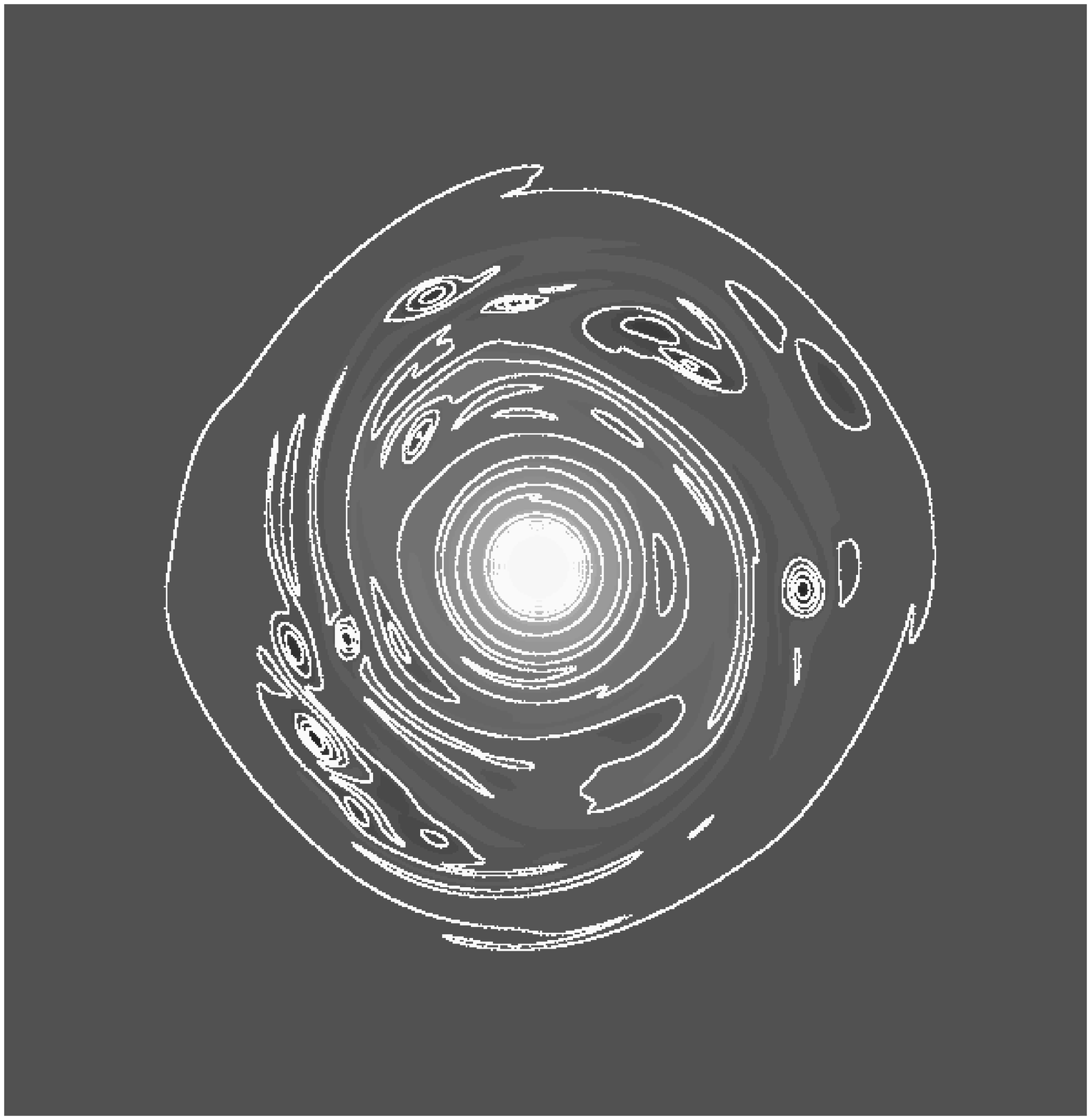,height=7cm}}
 \caption{Section of the vorticity field of a perturbed two--dimensional
    Keplerian disk, at time $t=0$ and $t=5$.
   }
\end{figure}


\begin{figure}

 \centerline{
 \psfig{figure=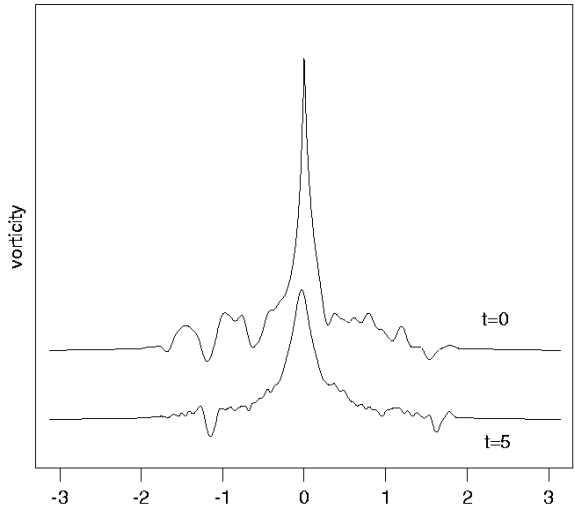,height=7cm}}
  \caption{Section of the vorticity field of a perturbed two--dimensional
     Keplerian disk, at time $t=0$ and $t=5$.
   }
\end{figure}

These results show that long-lived, coherent anticyclonic vortices
form in Keplerian shears, starting from an initially random perturbation
field. Several other simulations, including ones with different energies
in the Keplerian profile, different shapes of the initial perturbation,
and different ratios between the energy of the perturbation and that of
the unperturbed disk have given analogous results, provided that the
perturbation energy $\tilde E$ was larger than a certain threshold. For
the value of the
hperviscosity
$\nu_2= 5\cdot 10^{-8}$ used in the previous
simulation, the requirement is that $\tilde E/E_{kep}$ should be larger
than about 0.001 for sustained vortices to form.   The fate of nonlinear
perturbations did not, as might have been supposed, lead to disorganized
motions, but to coherent structures.  These are anticyclonic vortices
and they typically survive for tens or even hundreds of rotation
times before being dissipated.   Whether the final dissipation is a
numerical artifact or a real phenomenon is not decidable as yet and this
may be relevant to a discussion about whether the Keplerian profile
should be considered as nonlinearly unstable or not (Balbus and Hawley
1996; Dubrulle and Zahn 1991).
When we ran simulations with Newtonian viscosity, $p=1$, and $\nu_1=2\cdot
10^{-4}$ in eq.(\ref{dissip})), we found that coherent vortices form in
this case as well. The only difference is that the vortices obtained with
$p=1$ are slightly broader than those obtained with a $p=2$
hyperviscosity, as commonly observed for turbulence simulations
in a homogeneous background flow.

The size of the viscosity coefficient plays an interesting role.
The larger is the viscosity, the broader are the vortices. For large
values of $\nu_2$, small vortices cannot form and only a few of the
largest vortices survive. Conversely, the smaller is the viscosity, the
more numerous
the
small-scale vortices are. Simulations with
$\nu_2=5\cdot 10^{-9}$ show the presence of many
small-scale anticyclonic vortices that
penetrate well inside the central regions of the Keplerian disk.
The smaller is the viscosity, the smaller is the minimum energy ratio
$\tilde E/E_{kep}$ which is required in order to generate the
vortices.

The roles of both the amplitude of the initial disturbance and
of the viscosity emerge in the definition of two length scales that
are used in discussions of geostrophic turbulence.  In the present
context, a length relating to the shear, $L_S=\sqrt{U/\Omega'}$
may be defined where $\Omega' = d\Omega/dr$ and $U$ is a characteristic
initial velocity disturbance.  The shear length, $L_S$ is the disk
analogue of the Rhines scale of geostrophic turbulence theory (see
Cho and Polvani 1996a; 1996b).  In both
situations, where the rotation rate varies in one direction, it is found
that perturbations with scales greater than $L_S$ become highly
anisotropic.  It is believed that such structures do not participate in
the inverse cascade of energy and remain elongated.  Their decay
takes place through radiation of Rossby waves.

Distinct vortices can form only with sizes less than $L_S$.
On the other hand, vortices that are too small will be rapidly destroyed
by viscosity.  The viscous decay time of a structure of size $L$ is
of the order of $L^{2p}/\nu_p$.  For a vortex to live several rotation
periods, this time must be much larger than $\Omega^{-1}$, which requires
$L \gg L_\nu = (\nu_p/\Omega)^{1/2p}$.  Thus, we expect that the copious
formation of coherent vortices will be possible only if $L_\nu \ll
L_S$.  If the viscosity is too large, or the shear is too strong, $L_\nu$
and $L_S$ become comparable and vortices cannot form.  For $p=1$
and with $\Omega/\Omega'=r$, the criterion is that $Ur/\nu \gg 1$.
Though this is in a sense to be expected, the ingredients in the
derivation are somewhat different than they are in the usual derivation
of criteria involving the Reynolds number. Our Reynolds number is the
ratio of the $\beta$ term from the Coriolis force to the viscous term.

We have observed in these simulations that vortices form from general
perturbations having initial sizes of the order of $L_\nu$. The vortices
then merge with each other, and the size of the surviving vortices grows
with time.  In a homogeneous background, the only limit to this growth is
set by the size of the domain.  In a shear flow, the scale $L_S$ sets an
upper limit to the vortex size. Thus, the older is a vortex, the
closer its size is to $L_S$. At late evolutionary stages, most
anticyclonic vortices have a size comparable with $L_S$.
Since the strength of the local shear depends on the radius $r$,
and $L_S=L_S(r)$, the range of radii in which vortices form will
depend on the variation of $\nu$, hence density, with $r$.  In the
present simulations with constant $\nu_p$, we see larger vortices
farther out.

\section{Magnetic Effects}

The inclusion of a magnetic field in the disk can produce linear
instability and so change the fluid dynamics in an important way
(Balbus and Hawley 1998).  We
study the effect of a purely horizontal magnetic field on the disk
dynamics discussed in the previous section in the usual MHD approximation
with displacement current neglected.  The field is assumed to be not so
strong as to modify the thin layer assumptions.  We express it in terms
of a scalar magnetic potential $a({\bf r},t)$ such that the horizontal
field ${\bf B}=(B_x,B_y)$ is given by $B_x=-\partial a/\partial y$ and
$B_y=\partial a/\partial x$.  This magnetic field is horizontally
non-divergent so any vertical field that is present must be independent
of $z$.

The current is $\nabla \times {\bf B}$ and is in the vertical direction.
This does not fit in well with the thin layer image and is more suited
to the picture of two-dimensional incompressible flow.  We nevertheless
proceed formally to consider what happens when  the Lorentz force
${\bf j}\times {\bf B}$ is included in the dynamics of the previous
section, where the current ${\bf j}$ is purely in the $z$-direction and
its magnitude is given by
\begin{equation}
j=\nabla^2 a
\ .
\label{pot}
\end{equation}
The equation of motion for
the vertical vorticity becomes
\begin{equation}
{{\partial \zeta} \over {\partial t}}
+ [\psi,\zeta] - [a,j] = D_\zeta
\label{mhd1}
\end{equation}
where the various quantities are as before.  In addition, the induction
equation for the magnetic field, when expressed in terms of the
magnetic potential, is
\begin{equation}
{{\partial a} \over {\partial t}}
+ [\psi,a] = D_a
\ .
\label{mhd2}
\end{equation}
The Ohmic dissipation $D_a$ term is given by
\begin{equation}
D_a=(-1)^{q-1} \eta_q \nabla^{2q} a \ .
\end{equation}

In the absence of dissipation, eqs.(\ref{mhd1}) and (\ref{mhd2})
have three quadratic invariants: the total (kinetic
plus magnetic) energy $E= 1/2 \int (\nabla \psi)^2 dxdy +
1/2 \int (\nabla a)^2 dxdy$, the integral of the squared magnetic
potential, $A=\int a^2 dxdy$, and the cross helicity
$H=\int {\bf u}\cdot {\bf B} dxdy$. The simultaneous conservation
of these quantities induces a direct cascade of energy, from large
to small scales, and an inverse cascade of magnetic potential,
from small to large scales (Fyfe \& Montgomery 1976;
Fyfe {\it et al.} 1977). One thus expects to see
coherent structures in the current field rather than in the
vorticity. Since the three-dimensional MHD equations
have three quadratic invariants as well, the fact that
energy cascades to small scales, makes 2D MHD closer to the
full three-dimensional MHD problem than 2D Navier-Stokes is to
3D hydrodynamics.

The 2D MHD system has a wide range of dynamical behaviours.
In the limit of weak magnetic field, the magnetic potential
behaves as a passive scalar, and the term $[a,j]$ in
eq.(\ref{mhd1}) can be discarded. In this limit, one
recovers the behavior of the 2D Navier-Stokes equations
(\ref{2d}). An initially weak magnetic field does however grow with time,
and in later stages the dynamics is significantly affected by
magnetic forces.

An important diagnostic quantity is the global correlation
between the velocity and the magnetic fields defined as
$C=H/E$.  If the velocity and magnetic fields are nearly aligned,
the absolute value of the global correlation coefficient is close to one.
Several simulations have shown that, if $|C|$ is larger than about 0.2
initially, then it grows with time toward the value $|C|=1$. In this
process, called dynamic alignment, Lorentz forces cause the
magnetic and velocity fields to become parallel (or anti-parallel) to each
other. On the other hand, if $C \approx 0$ initially, then the two fields
remain uncorrelated.

Numerical simulations of the dissipative 2D MHD equations
with $D_\zeta \ne 0$ and $D_a \ne 0$, in a
homogeneous background reveal the
formation of current and vorticity sheets (Biskamp \& Welter 1989).
Coherent magnetic vortices also form and dominate the
current distribution and the dynamics of the system
(Kinney {\it et al.} 1995) so that, at later times, the
fluid vorticity plays a minor role, concentrating in sheets at
the periphery of the magnetic vortices.
In the individual magnetic vortices, there may be a non-zero
correlation between vorticity and current. The sign of this
correlation
is different in different vortices, leading to an
approximately zero global correlation between the magnetic and
the velocity fields.

The question that we consider here is whether magnetic vortices can
form in a shearing background. To address this issue, we have run a
series of simulations on the behavior of a magnetized, barotropic
Keplerian disk, by modifying the code described in the previous
section to deal with (\ref{mhd1})-(\ref{mhd2}).  In the runs discussed
below, we have chosen a hyperviscous dissipation with $p=q=2$,
$\nu_2=5\cdot 10^{-8}$ and $\eta_2=2.5\cdot 10^{-7}$. This gives a
magnetic Prandtl number $Pr_{mag}=\nu_2/\eta_2=0.2$. Runs with
$Pr_{mag}=1$ have provided analogous results.

In simulating a magnetized disk, we have to specify the initial
background current distribution.  Though it seems
likely that an initial field
will quickly develop a strong toroidal presence, we have decided to
leave the matter open, and so we show the results of two different
simulations. In the first, we assume no background current; here the
magnetic field is present only as a perturbation of a basically
unmagnetized background state. In the second type of simulation, we
assume the presence of a Keplerian background current, with the same
amplitude as the background vorticity field at the outset. In both cases,
the background vorticity is assumed to be Keplerian, as in
Section 3.

In the case with no background current, the background Keplerian vorticity
field has energy $E_{kep}=0.5$.
The typical rotation time at $r=\pi/4$ is
$T_{kep}(\pi/4)\approx 2$ and the simulation has been run
up to a time $T=20$.
In this simulation, we initially
perturb
the magnetic field
away from zero and do not perturb the vorticity;
the perturbation energy is
$\tilde E \approx 0.004 E_{kep}$. The spectrum of the initial
magnetic perturbation is
given by eq.(\ref{spectrum}).
This case is characterized by an almost zero correlation
between the initial magnetic and velocity fields.

Figures 4 and 5 show the vorticity and the current at time $t=10$ and
they reveal that strong magnetic vortices form in the
presence of an initial Keplerian shear. As already observed by
previous studies without a background shear, the
vorticity is dominated by magnetic effects, with fluid vorticity
concentrating in sheets and in weaker structures associated with
the magnetic vortices.  In the evolved fields, there is no global
correlation between vorticity and current.

\begin{figure}

  \centerline{ \psfig{figure=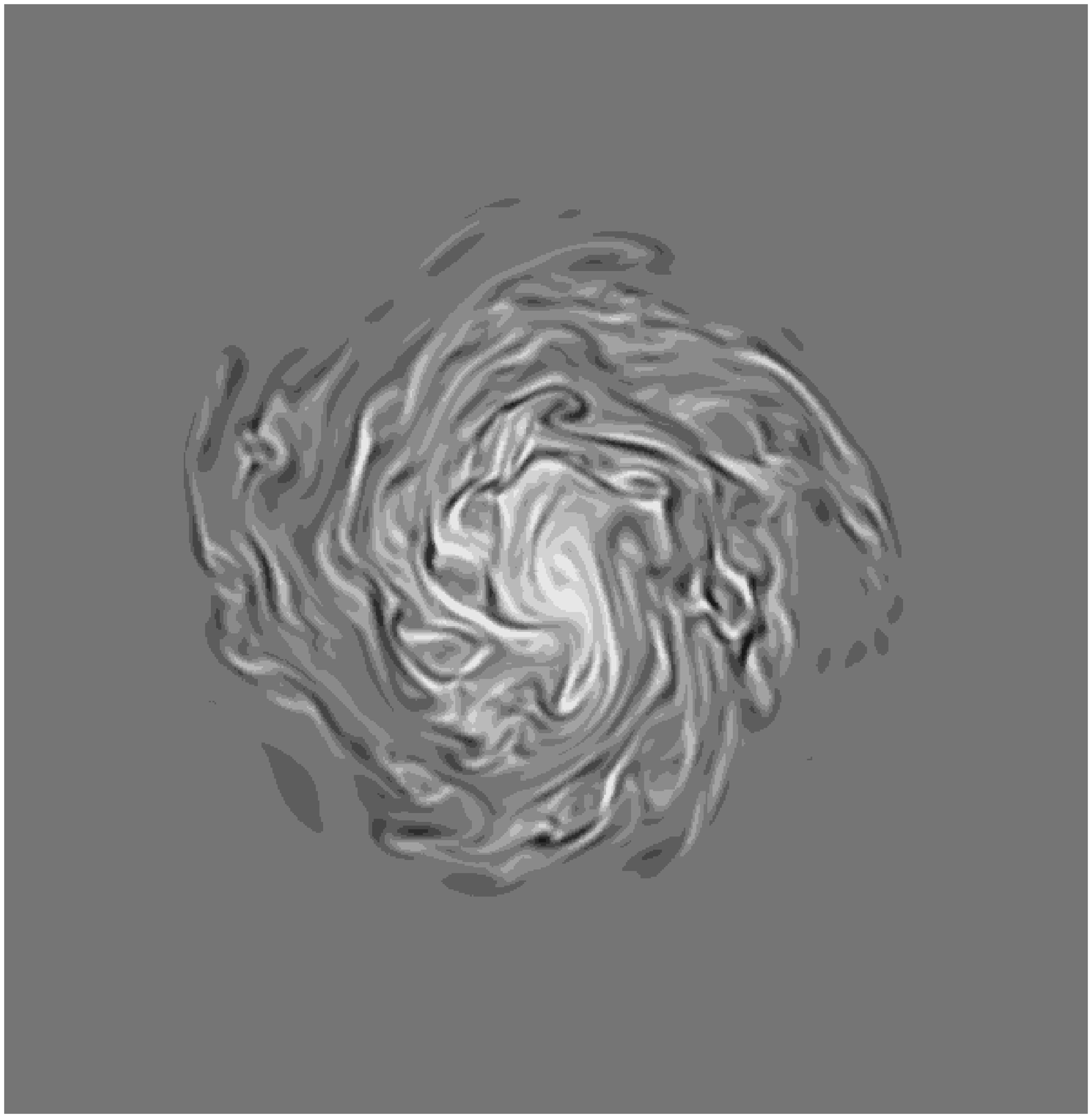,height=7cm} \psfig{figure=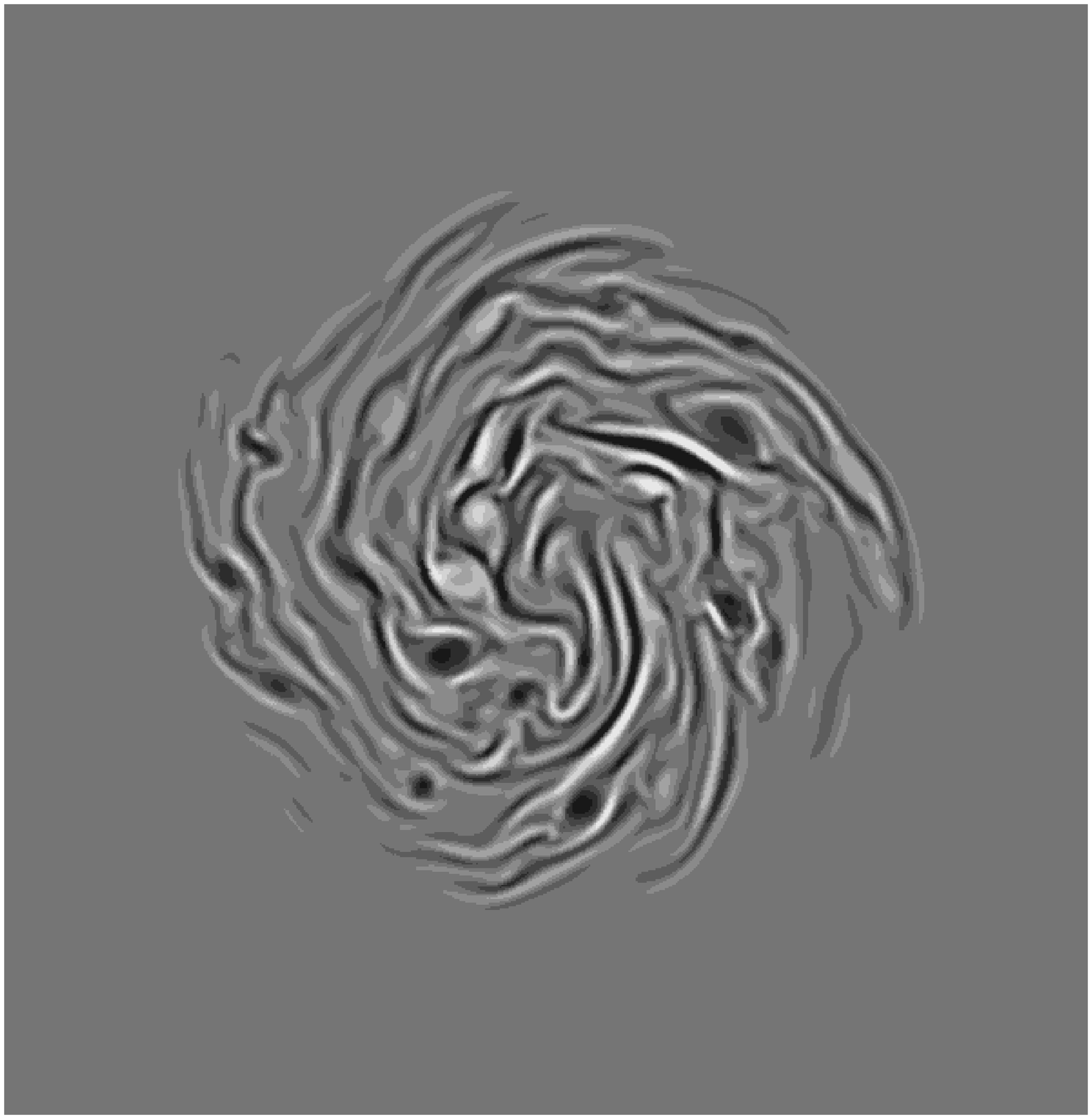,height=7cm}}
  \caption{Vorticity (figure 3a) and current (figure 3b) at time $t=10$
     for a magnetized, two--dimensional initially Keplerian disk with
     no initial background current.}
\end{figure}


A significant difference from the non-magnetic case is found in the
evolution of the mean vorticity profile. Without the magnetic field
the Keplerian profile is linearly stable to perturbations and
we find that the average radial profile remains Keplerian during the
evolution.  In the MHD simulation shown here, the presence of the magnetic
perturbation leads to a modification of the background Keplerian
profile, which becomes flatter as the evolution proceeds.
This is not surprising, as the magnetic torques enter the dynamics
significantly.

Figures 6 and 7 show the vorticity and
the current obtained when the current has a background Keplerian component
$J(r)$, equal in strength to the background vorticity profile $\Omega(r)$.
No vortices form in this case, the two fields becoming rapidly
fully correlated and frozen into alignment due to the presence of
an initial correlation $|C| > 0.2$.   The terms $[\psi,\zeta]$ and $[a,j]$
then cancel each other in eq.(\ref{mhd1}), and the field remains in its
Keplerian state, slowly decaying due to dissipation.

\begin{figure}

 \centerline{ \psfig{figure=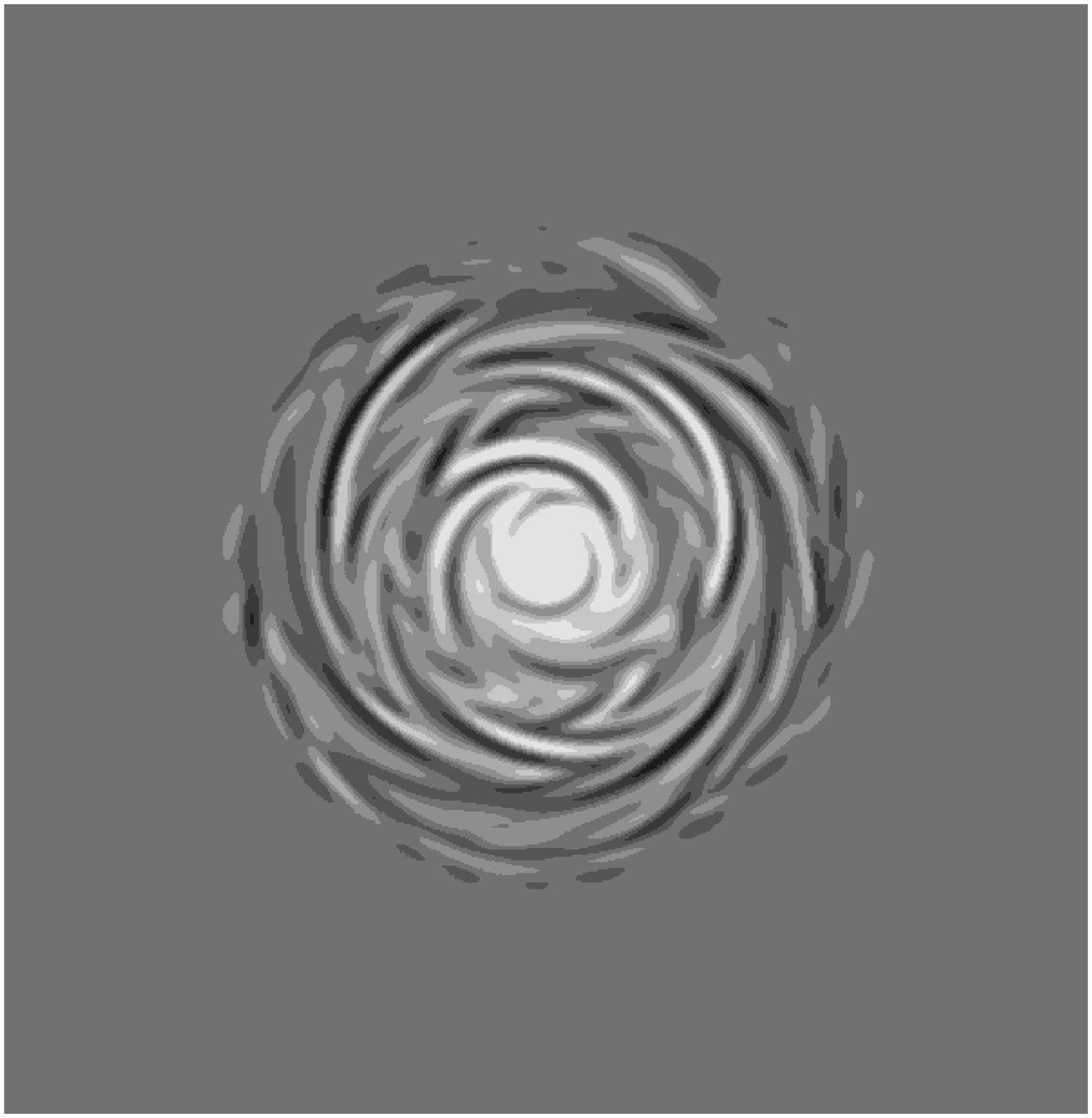,height=7cm}\psfig{figure=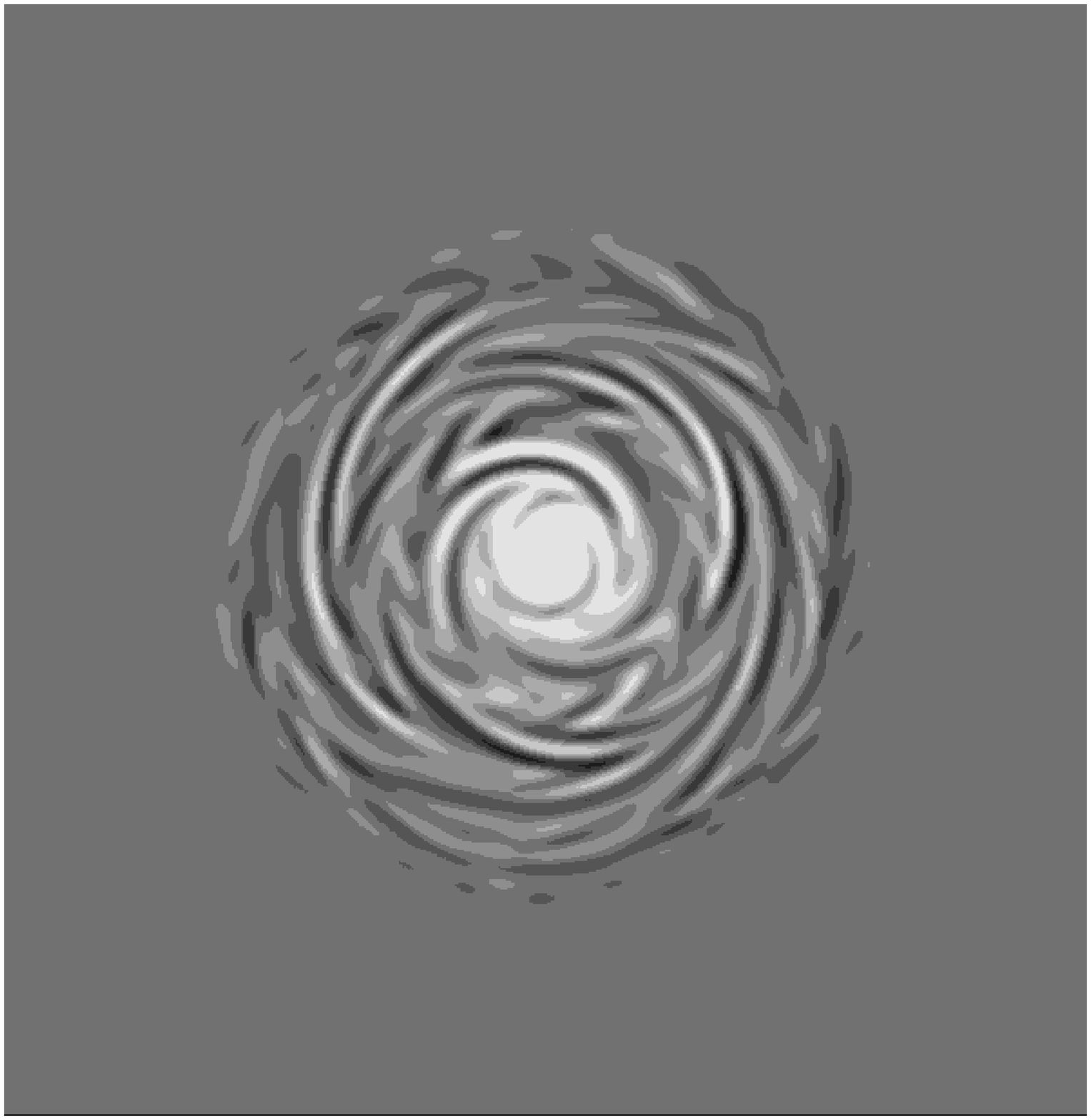,height=7cm}}
  \caption{Vorticity (figure 4a) and current (figure 4b) at time $t=10$
     for a magnetized, two--dimensional initially Keplerian disk with
     an initially Keplerian background current.}
\end{figure}


These simulations have thus shown that, when there is no background
Keplerian current, coherent magnetic vortices can be generated.
In the magnetic field, the strong cyclone-anticyclone asymmetry observed
for purely hydrodynamic disks is absent. The vorticity field is
dominated by the magnetic effects, and weaker vorticity
concentrations can be observed in the magnetic
vortices. In the absence of a background Keplerian current, the
vorticity profile does not remain Keplerian.  These
results are restricted to the unrealistic case of purely horizontal
magnetic field and do not include the influence of an external
gravitational force that might favor the maintenance of the Keplerian
background flow.  In the appendix we describe the inclusion
of this effect and observe that it does not change the general character
of the results.

\section{Disks With Variable Surface Density}

The full thin-layer theory is like that of a two-dimensional compressible
fluid, where density fluctuations are important, except that in the
thin-layer approximation these fluctuations propagate as compressible
gravity waves rather than acoustic waves
(Qian \& Spiegel 1994).
We present here some first
results from numerical simulations of the full thin-layer problem.
This problem is computationally more demanding than the version we have
just considered so that they involve even more modest parameter values
than the incompressible case.  We begin by recalling some of the
theoretical background for the thin-layer simulations
(Balmforth {\it et al.} 1992).
To do that, we temporarily ignore the viscous term in order to keep the
discussion simple; it is restored for the numerical computations.

In the full thin-layer theory of section 2, we solve the
continuity equation (\ref{surf}) together with the vertically integrated
counterpart of the momentum equation (\ref{NS}).  Since vertical motions
are negligible in thin layers, we have hydrostatic balance in the vertical
direction.  The simplified ${\hat z}$-component of the momentum equation
then integrates to:
\begin{equation}\label{swviw}
 \int_{-h}^h {1\over \rho}{\partial p\over\partial z} dz =
 \int_{-h}^h {\partial H\over\partial z} dz
   = - \int_{-h}^h {\partial \Phi\over\partial z} dz ,
\end{equation}

\noindent
where we have introduced the specific enthalpy $H$ according to the
definition provided by the first equality in (\ref{swviw}).
For a thin layer, the thickness of the disk, from $h$ above to $-h$ below
the plane, is small compared to the radial coordinate, $r$, so
this equation tells us that the enthalpy is well approximated by
\begin{equation}
H = {GM\over 2 r^3} \left(h^2 - z^2 \right) \ .
\label{enth}
\end{equation}

Since the pressure, and therefore the enthalpy, vanishes at
$z=\pm h$, (\ref{swviw}) and (\ref{enth}) become

\begin{equation}\label{eta}
\eta(x,y,t) = H(x,y,0,t) = {GM h^2 \over 2 r^3} ,
\end{equation}

\noindent
where $\eta$ is the specific enthalpy to be used in the thin layer.
Then, an effective equation of state for the layer can be found by
substituting the polytropic formula for enthalpy, $H = K {\gamma\over
\gamma -1} \rho^{\gamma-1}$, into  (\ref{swviw}), allowing the density to
be expressed as a function of the disk thickness; we get
\begin{equation}
\rho = \left[ {\gamma -1\over \gamma K}
        {GM\over 2 r^3}\left(h^2-z^2\right)
\right]^{1/\gamma -1}.
\label{eqn}
\end{equation}
This relation, when inserted into the definition of the surface
density, equation (\ref{sigma}), yields the vertically integrated
equation of state,
\begin{equation}\label{sw2}
\sigma = \sigma_0 \left[\eta^{\gamma +1\over\gamma-1} r^3\right]^{1/2} ,
\end{equation}

\noindent
where $\sigma_0$ is a constant.  Because of the dependence on
the local vertical gravity, the equation of state depends on $r$.

The horizontal momentum equation is now
\begin{equation}\label{swmom}
{{D {\bf u}}\over {Dt}}=-\nabla\left( \eta + \phi \right)
 +{\bf D}_{\bf u}
\end{equation}

\noindent
where $u$ is the two-dimensional velocity ${\bf u} = (u, v)$, $\phi$
is the gravitational potential in the plane
($\phi(x,y) = \Phi(x,y,0)$)
and the viscous term has been restored.  The expression for the potential
vorticity of the disk has
been given in equation (\ref{pote}).

We have solved equations (\ref{surf}), (\ref{sw2}) and (\ref{swmom})
in finite-difference form, on grids of $256\times 256$ and $512\times
512$ using a modified form of the Miami Isopycnal Coordinate Model
(MICOM),  a second order algorithm which
incorporates a flux-corrected transport of the surface
density.  This code has been used successfully for a wide variety of
oceanic flows; in particular, it has been used to study the ocean
thermocline, another thin layer whose equations are like the ones
studied here.  The special feature of this code, not contained in
typical shallow fluid codes, is that it allows for a vanishing
layer thickness (Sun {\it et al.} 1993).

For the initial conditions, we need a background disk.  For this we
used steady axisymmetric models with angular velocity profile given
by
\begin{equation}
\Omega
  = \left({1\over r}{d{\bar \eta}\over dr}+{1\over r}
   {d\phi\over dr}\right)^{1/2} \; ,
\label{sw1}
\end{equation}

\noindent
where $\bar \eta$ is the time-independent enthalpy. When (as is
usual in the mean state) the enthalpy variations are negligible, we
recover from (\ref{sw1}) the Keplerian velocity $\Omega_K =
r^{-3/2}$ for the potential $\phi =-1/r$.

Since enthalpy variations are not negligible, the variations of the
background surface density
lead to both a departure from
Keplerian flow, according to (\ref{sw1}), and to a modulation of the
background potential vorticity, according to (\ref{pote}) in a way that is
analogous to topographic effects frequently encountered in geophysical
problems.

To simplify the application of boundary conditions, we have taken
the case of an annulus with enthalpy distribution

\begin{equation}\label{sw3}
\bar{\eta}(r) = \eta_0 (r-r_i)^{n_i} (r_o-r)^{n_o} ,
\end{equation}

\noindent
where the subscripts $i$ and $o$ mean inner and outer.  This
$\bar{\eta}$ vanishes at two radii ($r=r_i$ and $r=r_o$), always
chosen to be well within the computational domain. Since we
apply boundary conditions only on the edges of the
domain, this makes for an insensitivity to the boundary conditions.

The computational domain extends from $-2$ to $2$ and we have used
typical values for the inner and outer disk radii of
$1/4$ and $7/4$, respectively.  The computations were
performed in a frame co-rotating with
the radius $r=1$.  The angular velocities do not depart
significantly from Keplerian values; at $r=1/3$ the
orbital period is $
\approx
1$.

In figure 8 we display the enthalpy for three steady disks, having
$n_i = {3\over 4}$ and $n_o = {3\over 4}, {4\over 4}, {5\over 4}$.
The corresponding potential vorticity is also plotted (middle),
but since $q$ is not well defined where $\sigma$
vanishes, we plot only those regions where the surface density exceeds
some minimum value: $\sigma > \sigma_0$.  The singular nature of $q$
indicates the presence of a boundary layer at the disk edges.  At
present we are interested only in the behavior far from such
boundary layers, so we can choose to represent
the inverse of $q$, or what is called the {\sl potential thickness}
(figure 8, bottom), rather than the potential vorticity.

\begin{figure}
 \centerline{ \psfig{figure=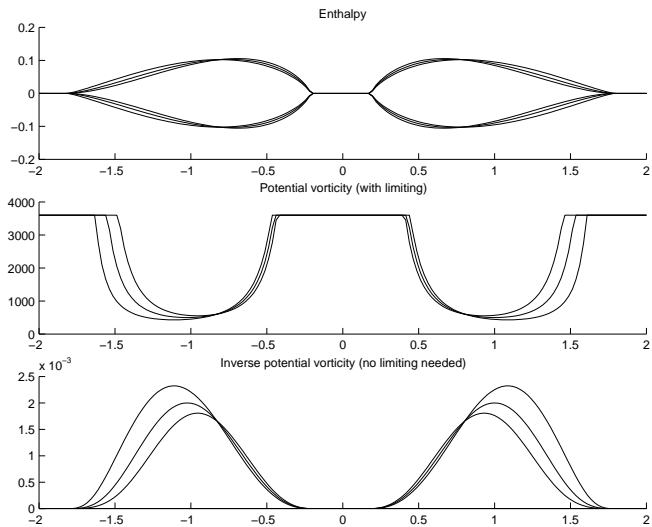,height=7cm}}
  \caption{Enthalpy (top), potential vorticity (middle), and
    potential thickness (bottom) for three thin disks having
    respectively $n_o={3/4}$, ${4/4}$ and ${5/4}$.
    }
\end{figure}

In shallow fluids, gravitational forces will act to restore perturbations
made to the layer thickness.  But in the presence of rotation, this
process excites local shears in the flow.  Large perturbations will be
deformed in this process since there is a scale above which rotation
dominates vertical stability.  In geophysics this critical scale is
called the (Rossby) deformation radius, $L_D$
(Pedlosky 1987),
given by $L_D^2 = 2 \bar \eta r/ f^2 h$.  Vortices larger than the Rossby
radius have quiet centers encircled by rings of rapidly varying potential
vorticity.  This manifestation of $L_D$ has been revealed by numerous
laboratory and computational results, which show how $L_D$ constrains the
extent of the perturbing shear layers within a vortex.

To study the action of potential vorticity and the behavior of
potential vortices in disks, we perturb the steady solutions given above.
One way to introduce small disturbances of $q$ is to simply perturb the
thickness. To leading order, the perturbed potential vorticity can be
expressed as
\begin{equation}\label{sw4}
{q} = \bar{q} - {\sigma ' \over \bar{\sigma}}\bar{q}
 + {\zeta ' \over \bar{\sigma}} ,
\end{equation}
where the last term is absent if we do not perturb
the absolute vorticity.  Using this expression, we build a
random distribution of perturbed potential vorticity, whose energy peaks
at a wavelength $\lambda \sim R_{disk}/10$, and superpose this
onto the steady disk.

The evolution of the initial disk given by (\ref{sw1}) and (\ref{sw3})
leads to an approximately steady disk with some negligible migration
of surface density toward $r=0$ as a result of viscous stress.
When the initial condition includes a perturbed potential vorticity
as described above and in (\ref{sw4}), the evolution is dominated by
the interaction of compressible vortices with each other and with the
background shear.  The disk at time $t=2$ is shown in figure 9.

\begin{figure}

 \centerline{ \psfig{figure=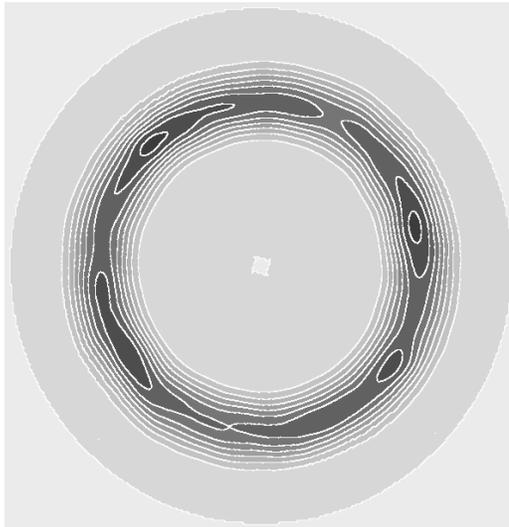,height=7cm}}
  \caption{Potential thickness (inverse potential vorticity)
    at time
    $t=2$ for
    a perturbed thin disk with $n_i=3/4$ and $n_o=5/4$.
    }
\end{figure}

The most obvious features found in the potential vorticity are:
(1) vortices which are too small are quickly eliminated by the
combined actions of the shear and dissipation;
(2) small potential vortices, if they are above this scale,
rapidly coalesce to form larger, more coherent
features; (3) cyclonic fluctuations tend to be easily
absorbed into the background shear of the disk;
(4) the more robust anticyclones can persist for many disk rotations.

\section{Conclusions}
At first sight it might seem that strong shears provide an abundant
source of angular momentum that would feed vortex production and
that would tightly wind local magnetic anomalies.  However, shears
also work against the conventional vortex formation process by
making the environment highly anisotropic.  In the context of the
Jovian atmosphere, shears also occur and the work on that situation
suggests that vortices cannot form if they are too large (Cho and
Polvani 1996a; 1996b).  Instead,
large structures give rise to bands in the jovian atmosphere.  We may
expect a similar behavior in disks and some of our shallow water
calculations have revealed large scale vortical structures with banded,
spiral character (Yecko 1995).  However, the calculations reported here
show that when the local disturbances are not so large, vortices
form in disks and that similar coherent structures form in mildly
magnetized disks.

In order to form these coherent structures, we need to have initial
perturbations of finite amplitude.  There is no doubt that these are
available in magnetized disks since they are linearly unstable.
The case of nonmagnetized disks is apparently in dispute.  The systems
we have studied are not unstable within the numerical schemes we have
been using.  Rather, they are excitable.  This means that a finite
amplitude disturbance with amplitude above a given threshold goes through
an interesting evolution (in fact formation of vortices) before the system
returns to its initial Keplerian state.  
If there were no magnetic fields, we would
need to provide a source of perturbations to produce vortices.  These
could be present even in nonturbulent disks around black holes since
they may be buffeted continually by incoming objects.  On the other
hand, magnetized disks will certainly have a continual source of
perturbations.

In any case, our numerical simulations of two-dimensional, barotropic
disks and of thin-layer, compressible disks have shown that sufficiently
energetic vorticity perturbations evolve into long-lived,
coherent vortices that may heavily affect the disk dynamics
in several ways before they decay.  In the presence of magnetic effects,
coherent magnetic vortices are formed instead of purely kinetic
vorticity structures.  These will likely give rise to a good deal of
variability and other phenomena seen on stars with similar activity.

\begin{acknowledgments}
We are grateful to Neil Balmforth and Steve Meacham
for discussion and comments. AP is grateful to JILA for
hospitality and support during much of his work on this report.
\end{acknowledgments}

\appendix
\section{Forcing Keplerian Flows}
In the previous sections we have discussed the evolution of a free
Keplerian disk, i.e., of a freely-decaying, initially perturbed
Keplerian profile. In the pure hydrodynamic case, the average profile
remains Keplerian, except for the fact that it is slowly dissipated by
the small hyperviscosity. In the MHD case, the average vorticity profile
rapidly becomes non-Keplerian.

On the other hand, we may suppose that external forces, which are
not described by the simple evolution equations used above, try
to mantain a background Keplerian rotation. These could be, for
example, the gravity of the central mass and the slow infall of
material at the edge of the disk. In this section, we study
the evolution of a perturbation of a Keplerian disk, whose average
profile is forced
to remain
Keplerian.

Suppose
that the vorticity
field is the sum of two terms, a constant Keplerian background
$\Omega(r)$, imposed from the outside,
and a perturbation $\tilde \zeta ({\bf r},t)$ which is free to evolve.
Dissipation
is assumed to act only on the perturbation field $\tilde \zeta$.
For a purely hydrodynamic disk, the
equations of motion become in this case
\begin{equation}
{{\partial \tilde \zeta} \over {\partial t}}
+ [\tilde \psi,\tilde \zeta] =\tilde D_\zeta\
-[\Psi,\tilde \zeta] - [\tilde \psi,\Omega]
\label{forced}
\end{equation}
where $\Omega(r)$ is the constant Keplerian profile,
$\Psi(r)$ is the corresponding stream-function, $\tilde
\zeta$ and $\tilde \psi$ are the vorticity perturbation and
perturbation stream-function, and $\tilde D_\zeta=
(-1)^{p-1} \nu_p \nabla^{2p} \tilde \zeta$ is the dissipation.
The Keplerian profile is a solution to the equation $[\Psi,\Omega]=0$,
this term is thus absent from eq.(\ref{forced}).

The term $F=
-[\Psi,\tilde \zeta] - [\tilde \psi,\Omega]$ represents an
external forcing on the evolution of the perturbation. If the
perturbation were capable of extracting energy from the overall
rotation of the disk, then it could grow and eventually stabilize,
without being necessarily dissipated. From another point of view,
this approach allows for studying the stability of a Keplerian
disk, by following the evolution of an initial perturbation. This
is just a special case of the general study of the stability of
(barotropic) shear flows. Numerical simulation of eq.(\ref{forced})
has shown that anticyclonic vortices form
in this case as well, notwithstanding the continuous forcing.
The perturbation does, however, decay with time
and simulations
run with other values of the disk parameters
---
viscosity, ratio of the perturbation energy to the
energy of the background Keplerian flow,
{\it etc.} ---
have led to analogous results.

In the MHD case,
the magnetic field acts as a catalyst that allows the vorticity
perturbation to extract energy from the Keplerian shear. For a
forced MHD disk the equations are
\begin{equation}
{{\partial \tilde \zeta} \over {\partial t}}
+ [\tilde \psi,\tilde \zeta] -[\tilde a, \tilde j] =\tilde D_\zeta
-[\Psi,\tilde \zeta] - [\tilde \psi,\Omega]
\label{formhd1}
\end{equation}
\begin{equation}
{{\partial \tilde a} \over {\partial t}}
+ [\tilde \psi,a] =\tilde D_j
\label{formhd2}
\end{equation}
where $\tilde a$ and $\tilde j=\nabla^2 \tilde a$ are the
perturbation on the magnetic potential and the current and
$\tilde D_j=(-1)^{q-1} \eta_q \nabla^{2q} j$. We have supposed
no background Keplerian current and the forcing is acting
only in the vorticity equation.

The numerical simulation of eqs.(\ref{formhd1},\ref{formhd2}) has shown
that magnetic
vortices do form also in the presence of a continuous
forcing. In this case, however, the perturbation energy grows
with time, and the Keplerian MHD disk is unstable. The presence of
the magnetic fields acts as a trigger for the instability, and
the kinetic energy of the perturbation grows with time.
Notwithstanding the Keplerian forcing on vorticity, also in
this case the background vorticity profile does not remain
Keplerian.

\end{document}